# Worldwide AI Ethics: a review of *200* guidelines and recommendations for AI governance

AI Robotics Ethics Society

**Nicholas Kluge Corrêa**[1]    **Camila Galvão**[2]    **James William Santos**[3]    **Carolina Del Pino**[4]

**Edson Pontes Pinto**[5]    **Camila Barbosa**[6]    **Diogo Massmann**[7]    **Rodrigo Mambrini**[8]    **Luiza Galvão**[9]

**Edmund Terem**[10]    **Nythamar de Oliveira**[11]

February 19, 2024

## Abstract

The utilization of artificial intelligence (AI) applications has experienced tremendous growth in recent years, bringing forth numerous benefits and conveniences. However, this expansion has also provoked ethical concerns, such as privacy breaches, algorithmic discrimination, security and reliability issues, transparency, and other unintended consequences. To determine whether a global consensus exists regarding the ethical principles that should govern AI applications and to contribute to the formation of future regulations, this paper conducts a meta-analysis of 200 governance policies and ethical guidelines for AI usage published by public bodies, academic institutions, private companies, and civil society organizations worldwide. We identified at least 17 resonating principles prevalent in the policies and guidelines of our dataset, released as an open-source database and tool. We present the limitations of performing a global scale analysis study paired with a critical analysis of our findings, presenting areas of consensus that should be incorporated into future regulatory efforts. All components tied to this work can be found in this URL.

[1]Graduate Program in Philosophy. Pontifical Catholic University of Rio Grande do Sul/ University of Bonn. Porto Alegre/Bonn, Rio Grande do Sul/North Rhine-Westphalia, Brazil/Germany. nicholas@airespucrs.org.

[2]Graduate Program in Law. Pontifical Catholic University of Rio Grande do Sul. Porto Alegre, Rio Grande do Sul, Brazil. camila@galvaoadvogados.net.

[3]Graduate Program in Philosophy. Pontifical Catholic University of Rio Grande do Sul. Porto Alegre, Rio Grande do Sul, Brazil. james.santos@edu.pucrs.br.

[4]Psychology Undergraduate Program. Pontifical Catholic University of Rio Grande do Sul. Porto Alegre, Rio Grande do Sul, Brazil. c.pino@edu.pucrs.br.

[5]Graduate Program in Law. Pontifical Catholic University of Rio Grande do Sul. Porto Alegre, Rio Grande do Sul, Brazil. edson.pinto@fcr.edu.br.

[6]Graduate Program in Philosophy. Pontifical Catholic University of Rio Grande do Sul. Porto Alegre, Rio Grande do Sul, Brazil. camila.palhares@acad.pucrs.br.

[7]Graduate Program in Philosophy. Pontifical Catholic University of Rio Grande do Sul. Porto Alegre, Rio Grande do Sul, Brazil. diogo.massmann@edu.pucrs.br.

[8]Graduate Program in Law. Pontifical Catholic University of Rio Grande do Sul. Porto Alegre, Rio Grande do Sul, Brazil. rodrigo.barbosa92@edu.pucrs.br.

[9]Law Undergraduate Program. Pontifical Catholic University of Rio Grande do Sul. Porto Alegre, Rio Grande do Sul, Brazil. luiza.galvao@edu.pucrs.br.

[10]Graduate Program in Philosophy. University of Johannesburg. Porto Alegre, Rio Grande do Sul, Brazil. 220050769@student.uj.ac.za.

[11]Graduate Program in Philosophy. Pontifical Catholic University of Rio Grande do Sul. Porto Alegre, Rio Grande do Sul, Brazil. nythamar.oliveira@pucrs.br.

## 1   Introduction

Since the end of our last "*AI winter,*" 1987 – 1993, AI research and its industry have seen massive growth, either in developed technologies, investment, media attention, or new tasks that autonomous systems are nowadays able to perform. By looking at the history of submissions in ArXiv (between 2009 and 2021),[12] starting from 2018, Computer Science-related papers have been the most common sort of submitted material, increasing tenfold. Also, when we examine the category of Computer Science alone, "*Computer Vision and Pattern Recognition,*" "*Machine Learning,*" and "*Computation and Language*" are the most submitted types of sub-categories. Note that these are areas where Machine Learning is the current paradigm.

Besides the number of papers produced, we have never had more capital invested in AI-related companies and startups, either by governments or Venture Capital firms (more than 90 billion USD in 2021 in the USA alone) and AI-related patents being registered [72].

With the expansion of the AI field/industry also came another, the "*AI Ethics boom,*" where a never-before-seen demand for regulation and normative guidance of these technologies has been put forward. The growth of AI has brought the risks and side effects of its use. The implications of unregulated AI translate into ethical concerns, for example, about privacy and surveillance, prejudice, and discrimination that may be more socially harmful than economically beneficial. Thus, philosophical concerns about the political and moral implications of the interaction between technologies and human judgment are urgent. Such a scenario has led several entities and companies worldwide to publish manifestos about their ethical principles concerning AI, 200 of which were mapped and analyzed by this meta-analysis.

Despite the number of publications on the subject, we can see the difficulty that countries have faced in defining guiding principles and rules that will regulate artificial intelligence due to the complexity of interests involved, as can be seen in discussions about the AI ACT in the European Union or even in the parliament of our country of origin, which discusses the Legal Framework for Artificial Intelligence.

That said, inspired by the work done by previous meta-analysts in the field, this study presents a systematic review of 200 documents related to AI ethics and governance to map whether there is any consensus or similarity among the ethical principles advocated by the institutions involved in the topic (industry, academia, civil society, etc.) that can or should be protected by future legislation.

Through this analysis, the reader will be able to see that there is a need for regulation and that one of the biggest challenges we face in the field today is that ethical principles cannot be universalized, making the standardization of contextual ethical parameters a real challenge in the search for regulation. Problems involving ethical boundaries and artificial intelligence systems are emerging and demanding responses from government and private companies. However, these responses do not account for the scope of what is yet to come. Laws and regulations are still principled at best and do not focus on restricting the development of these systems.

It was also possible to notice that the highest number of documents manifesting ethical concerns come from "AI-developing" countries, while nations that are "AI-users" do not have so many manifestos. Another perceived point is that most of these documents are superficial and generic concerning their practical application and have a non-binding character, which hinders their effectiveness.

The exploration of these and other points is the aim of this study.

## 2   Related Work

One of the first studies to promote a meta-analysis of published AI ethical guidelines was that of Jobin et al. [31]. In this study, the authors sought to investigate whether a global agreement on emerging questions related to AI ethics and governance would arise. The research identified 84 documents containing ethical guidelines for intelligent autonomous systems, some of them being one of the most cited guidelines in the literature, like the Organization for Economic Co-operation and Development (OECD) Recommendation of the Council on Artificial Intelligence [69], the High-Level Expert Group on AI Ethics Guidelines for Trustworthy AI [46], the University of Montréal Declaration for responsible development of artificial intelligence [44], the Villani Mission's French National Strategy for AI [36], among many others.[13]

---

[12]ArXiv in Numbers 2021. https://arxiv.org/about/reports/2021_usage.

[13]Jobin et al. [31] sample also contained documents from governmental organizations (e.g., Australian Government Department of Industry Innovation and Science [42]), private companies (e.g., SAP [55], Telefonica [53], IBM [10]), non-governmental organizations (e.g., Future Advocacy [1], AI4People [2]), non-profit organizations (e.g., Internet Society [63], Future of Life Institute [43]), academic institutions (e.g., AI Now Institute [28]), professional associations (e.g., IEEE [52]), among other types of institutions.

In addition to listing the most common principles among the documents analyzed,[14] the authors also pointed to a *"substantive divergence concerning how these principles were defined"* (p. 1).[15] Furthermore, Jobin et al. work is careful not to impose any normative guidelines for the effectiveness of the mentioned principles. It attempts to raise the issue and map the global picture. However, the limited sample entails that their findings do not wholeheartedly address the obstacles of integrating ethical guidelines and AI development.

Another work that presented a similar type of analysis was that of Hagendorff [21]. Hagendorff's meta-analysis/critical review focused on a smaller sample (21 documents) but prioritized a specific set of documents. Documents classified as older than five years only referred to a national context,[16] not "specifically" related to AI (e.g., data ethics and the ethics of robotics), and corporate policies,[17] were not included in Hagendorff's study, which only chose documents that (according to his criteria) *"proven to be relevant in the international discourse"* (p. 3).

Even working with a smaller sample, Hagendorff's findings corroborate with those of Jobin et al., and the most cited principles found in his research were accountability (77%), privacy (77%), justice (77%), and transparency (68%). Jobin et al. and Hagendorff's studies also presented "blind spots" in the field of AI ethics, like the under-representation of ethical principles such as sustainability,[18] and the under-representation of documents produced by institutions from South America, Africa, and the Middle East. [19]

Other points cited by Hagendorff's work [21] served as motivation for the present meta-analysis, such as:

- The lack of attention given to questions related to labor rights, technological unemployment, the militarization of AI and the creation of LAWS (Lethal Autonomous Weapons Systems), the spread of disinformation, electoral interferences, misuse/dual-use of AI technology, some cited by last than half of the documents;

- The lack of gender diversity in the tech industry and AI ethics. According to Hagendorff's study, excluding the documents written by the research institute AI Now, a deliberately female-led organization, the proportion of female authors is only 31%;

- The short, brief, and minimalist views some documents give to normative principles. Some documents have a size of no more than 500 words;

- The lack of technical implementations for how to implement the defended principles in AI development (only 9% propose such implementations);

- The lack of discussion on long-term risks (e.g., AGI Safety, existential risks).

Hagendorf´s work on AI ethics deficits [22] and Jobin et al. [31] meta-analysis are valuable contributions to the field of AI ethics and provide the stepping stone to a worldwide analysis seeking the inclusion of a large number of other worth mentioned documents in several countries.

Finally, we would like to cite the work done by Fjeld et al. [15]. In their study, the authors worked with a sample of 36 documents originating from regions like Latin America, East/South Asia, the Middle East, North America, and Europe, being produced by a variety of institutions types, like governmental institutions (13 documents), private companies (8), professional associations and NGOs (5), intergovernmental organizations (2), and other types of multi-sectoral

---

According to the authors, most of their sample came from private institutions (22,6%), governmental organizations (21,4%), and academic research institutions (10,7%), other documents having varied origins (e.g., NGOs, non-profits, professional associations).

[14]In their sample [31], the detected ethical principles were Transparency, Justice/Equity, Non-maleficence, Accountability, Privacy, Beneficence, Freedom & Autonomy, Trust, Dignity, Sustainability, and Solidarity. Of these 11 ethical principles cited, five were the most recurrent: Transparency (86%), Justice (81%), Non-maleficence (71%), Responsibility (71%), and Privacy (56%).

[15]The top creators of these guidelines are European Union nations (especially France, Germany, and Italy), followed by the United States of America, the United Kingdom, Canada, Finland, Sweden, Denmark, Japan, China, India, Mexico, Australia, and New Zealand. In summary, the main actors from the Global North.

[16]However, Hagendorff maintained samples from the European Union, the USA, and China. In Hagendorff's words, AI *"superpowers"* (p. 3).

[17]Concerning these exclusion criteria, Hagendorff's allowed corporate policies of specific institutions (IEEE, Google, Microsoft, IBM) due to their *"well-known media coverage"* (p. 3).

[18]Only 16% of the documents reviewed by Jobin et al. cite this principle. In Hangerdoff's study, only 4% of the documents cite this principle.

[19]According to the NGO AlgorithmWatch, its AI Ethics Guidelines Global Inventory contains 167 documents. None of these documents predate the year 2013, and only two have their origin tied to Southern Africa and Southern Asia (*no documents produced by Latin America are listed*).

initiatives (7).[20] Like in the study of Jobin et al. [31], Fjeld et al. [31] also cite the variability in how such principles are defined.[21]

However, once again, the restrictions of the sample analyzed can be seen since it also focused on documents that specifically address Autonomous and Intelligent systems (A/IS) (as defined by the IEEE [41]), leaving robotics and other AI-tangent fields aside (e.g., Data Science). Documents that did not present normative directions (i.e., only descriptive samples) or addressed only specific AI applications (e.g., facial recognition) were filtered out.

Fjeld et al. also pointed out that their choice of a thirty-six documents sample size aimed to facilitate (besides all previously mentioned exclusion criteria) a data visualization framework that could deliver a side-by-side comparison of individual documents (p. 14).

Some of the conclusions and recommendations for future work that the authors arrive at are [15]:

1. Ethical principles can only be interpreted inside a specific cultural context and need to be embedded inside a stricter form of policy governance to have true normative strength. Thus, their effectiveness in international settings remains, at best, uncertain;

2. There is a gap between established principles and their actual application. Many documents only prescribe normative claims without the means to achieve them, while the effectiveness of more practical methodologies, in the majority of cases, remain extra empirical (i.e., they have not been sufficiently tested yet);

3. Convergence among principles does not mean that they are unanimous. Many of these principles have varied and even opposite descriptions. Also, other values and principles may still be hidden under this varying normative discourse;

4. Removing Data Science and Robotics from the scope of analysis hides the relationship between these fields with AI. Machine Learning is strongly related to data science, which cannot be forgotten as a part of the life cycle [64] of these systems (e.g., many issues related to Privacy occur during the "*Data Science*" part of a model creation). Lethal Autonomous Weapons Systems (LAWS) are of great concern in AI ethics, and we cannot forget that LAWS are robots (e.g., combat drones). Most AI systems are software. Demanding that software engineers be responsible during the development of AI systems is nothing more than ethics in software development/engineering;

5. Only a minority of the (governmental) documents adopt stricter forms of regulation (e.g., restrictions and prohibitions);

6. A more detailed study on how principles definitions vary, including a more diverse mapping of documents, would also provide more compelling results in future work;

7. It would be important to track how the documents relate to what is decided by courts, parliaments, and administrative bodies in the countries that produced/adopted them.

Fjeld et al. brush on specific points that illustrate the gap between normative claims and the means to achieve them, the contextual nature of the hermeneutics of ethical principles, and its reflexes on the disparities of descriptors. Nevertheless, the Fjeld et al. sample (despite being able to capture some sense of ethnic and geographical diversity) does focus on said "*prominent*" documents on AI ethics, making a forward-looking landscape of AI ethics far too incipient. To summarize, we can contribute by highlighting the issue of how to make ethical guidelines compatible with AI development, expanding this debate to more parts of the globe and more AI-related fields.

---

[20]According to the authors, eight principles were the most commonly cited in their sample: Fairness/Non-discrimination (present in 100% of the analyzed documents), Privacy (97%), Accountability (97%), Transparency/Explainability (94%), Safety/Security (81%), Professional Responsibility (78%), Human Control of Technology (69%), and Promotion of Human Values (69%). They also defined these main principles as "*themes*," each theme containing a different group and distribution of principles that align with each other (e.g., Promotion of Human Values: Human Values and Human Flourishing, Access to Technology, Leveraged to Benefit Society). The authors cite the defense of Human Rights as a recurring theme in these documents. However, and maybe even surprisingly, the institutions that bring forth the cause of human rights with more frequency are not governmental (46%) or intergovernmental organizations (67%), but private institutions (88%) and civil societies (NGOs/non-profits) (80%), which by their very nature lack the power to make their normative claims more than mere suggestions or recommendations.

[21]For example, in the 2018 version of the Chinese Artificial Intelligence Standardization White Paper [29], the authors mention that AI can serve to obtain more information from the population, even beyond the data that has been consented to (i.e., *violation of informed consent would not undermine the principle of Privacy*), while the Indian National Strategy for Artificial Intelligence Discussion Paper (National Institution for Transforming India [16]) argues that its population must become massively aware so that they can effectively consent to the collection of personal information. Fjeld et al. [15] study present other more nuanced forms of how these principles diverge in definitions.

There is more meta-analytical work done in AI ethics that we will not cite in depth. For example, we could cite Zeng et al. [71] systematic review (which produced one of the repositories used in this research) or the Global AI Policy monitoring sustained by the Future of Life Institute (FLI). The studies cited are those that worked with the largest sample (i.e., nº of documents), while other studies show similar results. For a complete review of meta-analytical research on normative AI documents, we recommend the work done by Schiff et al. [57], which cites many other important works.

We believe that many of the later analyses, following the work of Jobin et al. [31], suffer from a small sample size. Even if the documents selected by all the before-mentioned studies are indeed the "*most relevant*," that does not mean other relevant issues are not raised in less-known documents. Perhaps better tools for implementing normative principles in AI systems design are present in those documents. To help the AI community create a better understanding of our AI ethics global landscape, in this study, we sought to analyze a larger sample size, also proposing more diverse categories and typologies for the documents found.

## 3  Methodology

From the gaps pointed out in the previous meta-analyses, in this study, we sought to combine:

1. A quantitatively larger and more diverse sample size, as done by Jobin et al. [31]. Our sample possesses 200 documents originating from 37 countries, spread over six continents, in six different languages;

2. Combined with a more granular typology of document types, as done by Hagendorff [21]. This typology allowed an analysis beyond the mere quantitative regarding the content of these documents.

3. Presented in an insightful data visualization framework. We believe the data presentation done by authors like Hagendorff and Fjeld et al. [15] was not "*user-friendly*" or clear. Something that we tried to overcome in our work.

We used as primary sources for our sample two public repositories, the "*AI Ethics Guidelines Global Inventory*," from AlgorithmWatch (AW)[22] [23] and the "*Linking Artificial Intelligence Principles*" (LAIP) Guidelines.[24] The AW repository contained 167 documents, while the LAIP repository contained 90 documents.

Initially, we checked for duplicate samples between both repositories. After disregarding duplicates, we also scavenge for more documents through web search engines and web scraping, utilizing keywords such as "*Artificial Intelligence Principles*," "*Artificial Intelligence Guidelines*," "*Artificial Intelligence Framework*," "*Artificial Intelligence Ethics*," "*Robotics Ethics*," "*Data Ethics*," "*Software Ethics*," "*Artificial Intelligence Code of Conduct*," among other related keywords. We limited our search to samples written/translated in one of the four languages our team could cope with: English, Portuguese, French, German, and Spanish.

It is important to remember that this is an incomplete sample. Due to the language barrier, our sample consisted of only documents available in the languages we had proficiency.

Diving into our methodological setting, we refer to "*guidelines*" as documents conceptualized as recommendations, policy frameworks, legal landmarks, codes of conduct, practical guides, tools, or AI principles for the use and development of this type of technology. Most of these documents present a form of Principlism, i.e., normative documents based on ethical principles. Even purely practical documents/tools have as a foundation certain ethical principles (e.g., debiasing tools built upon the principle of Fairness). These principles are a foundation for how AI technologies should be used and developed. From them, documents usually create normative mechanisms/instruments, e.g., codes of conduct, ethical frameworks, governance frameworks, software development tools, impact assessment tools, etc.

Now, deconstructing the expression "*AI Technologies*," with "*AI*," our scope of interest encompasses areas that inhabit the multidisciplinary umbrella that is Artificial Intelligence research, such as Statistical Learning, Data Science, Machine Learning (ML), Logic Programming/Symbolic AI, Optimization Theory, Robotics, Software Development/Engineering, etc.

And with the term "*Technologies*," we refer to specific tools/techniques (e.g., Rule-Based systems, Convolutional Neural Networks, Transformers), applications (e.g., image recognition software, chatbots, applied robotics), and services (e.g.,

---

[22]https://inventory.algorithmwatch.org/. .

[23]AlgorithmWatch launched in April 2019 its "*Inventory of AlgorithmWatch principles, voluntary commitments, and frameworks for ethical use of algorithms and AI.*" The institution clarifies that although its inventory is one of the most complete, it is still a mere slice of a much larger (and as yet little explored) landscape.

[24]https://www.linking-ai-principles.org/.

Netflix recommendation system, Banks automated fraud detection, etc.). The term refers to technologies used for automating decision processes and mimicking intelligent/expert behavior.[25]

We analyzed our sample (200 documents) in two phases. In phase one, a team of ten researchers was assigned different quotas of documents. Researchers were responsible for reading, translating when needed, and hand-coding pre-established features. The first features looked for:

- Institution responsible for producing the document;
- Country/World Region of the institution;
- Type of institution (e.g., Academic, non-profit, Governmental, etc.);
- Year of publication;
- Principles (principles were considered to be all those values/goals/commitments/rights defined/supported by the institutions/authors that listed them in their guidelines);
- Principles description (i.e., the words used in a document to define/support a given principle);
- Gender distribution among authors (inferred through a first name automated analysis);
- Size of the document (i.e., word count).

We did not break principles ("*themes*") into related sub-principles like in the Fjeld et al. [15] study but merely paraphrased all documents' definitions for the defended principles found. We used this approach to avoid creating a biased pool of sub-principles, a problem already cited by Fjeld et al. (p. 16), since the majority of documents in our sample, like in previous studies, originated in Western European and North American countries.

In the first phase, based on past works, we established a list of principles so our team could focus their search. These were: Accessibility, Accountability, Auditability, Beneficence/Non-Maleficence, Dignity, Diversity, Freedom/Autonomy, Human-Centeredness, Inclusion, Intellectual Property, Justice/Equity, Open source/Fair Competition, Privacy, Reliability, Solidarity, Sustainability, Transparency/Explainability. The first phase helped us refine our list of principles. We determined that similar principles could be aggregated under the same category (by expanding its name, e.g., Diversity/Inclusion/Pluralism/Accessibility)since they upheld resonant values and ideas (e.g., Inclusion, Diversity, Accessibility).

In the second phase of our analysis, we also set new principles to be contemplated. These principles were later added because:

1. We were initially unaware of their existence in the current debate.
2. They appear to be cited a sufficient number of times (>10).
3. They could not be integrated into another category without redefining it.

The definition of these categories was subjective, but a deep dive into our sample provided the input needed. To create our "*overall definition*" for each category, a text mining technique called n-gram analysis was utilized, where the successive repetition of words (and groups of words) was counted in every principle category (Fig. 1).[26]

The defined principles helped aggregate similar and resonating values while maintaining significant typological differences. Bellow, the reader can find the definition that we gave to each of them:

- *Accountability/Liability:* accountability refers to the idea that developers and deployers of AI technologies should be compliant with regulatory bodies, also meaning that such actors should be accountable for their actions and the impacts caused by their technologies;
- *Beneficence/Non-Maleficence:* beneficence and non-maleficence are concepts that come from bioethics and medical ethics. In AI ethics, these principles state that human welfare (and harm aversion) should be the goal of AI-empowered technologies. Sometimes, this principle is also tied to the idea of *Sustainability*, stating that AI should be beneficial not only to human civilization but to our natural environment and other living creatures;

---

[25]However, it is important to remember that "*intelligence*" is not a well-defined concept backed by a strong scientific consensus.
[26]The word frequency for every principle category was calculated using the "**sklearn.feature_extraction.text.CountVectorizer**". We also used a custom stopword list file to filter common words (like "*the*" or "*and*") found in our sample (like "*ai*" and "*intelligence*"). The code for our implementation can be found in this repository: https://github.com/Nkluge-correa/worldwide_AI-ethics.

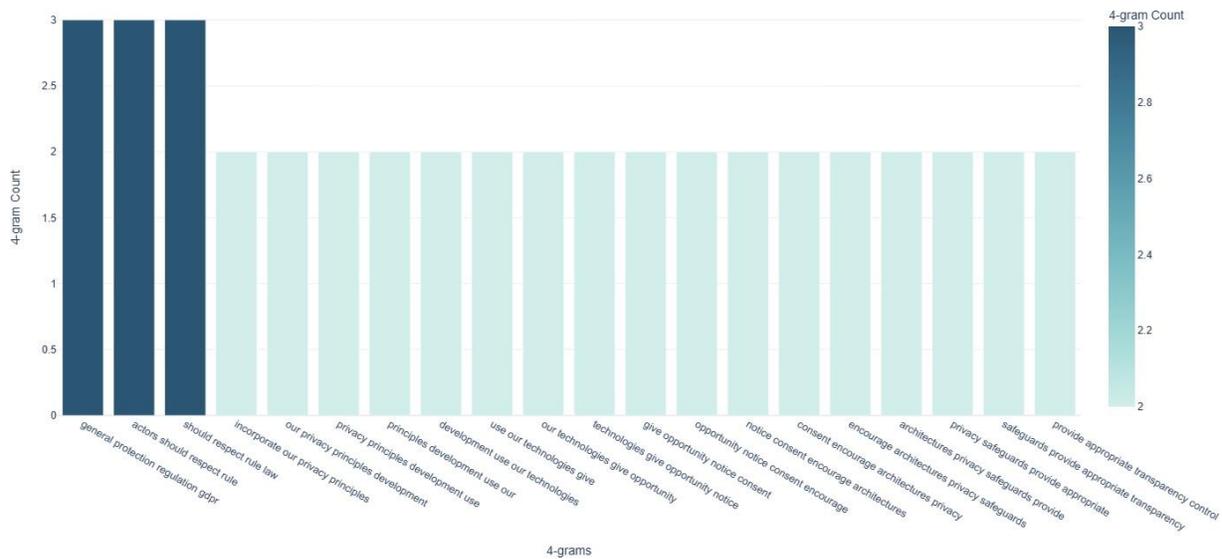

Figure 1: Top-20 recurrent four-grams for the 137 descriptions of "Privacy."

- *Children & Adolescents Rights:* the idea that the rights of children and adolescents must be protected. AI stakeholders should safeguard, respect, and be aware of the fragilities associated with young people;

- *Dignity/Human Rights:* this principle is based on the idea that all individuals deserve proper treatment and respect. In AI ethics, respect for human dignity is often tied to human rights (i.e., Universal Declaration of Human Rights);

- *Diversity/Inclusion/Pluralism/Accessibility:* this set of principles advocates the idea that the development and use of AI technologies should be done in an inclusive and accessible way, respecting the different ways that the human entity may come to express itself (gender, ethnicity, race, sexual orientation, disabilities, etc.). This principle is strongly tied to another set of principles: Justice/Equity/Fairness/Non-discrimination;

- *Freedom/Autonomy/Democratic Values/Technological Sovereignty:* this set of principles advocates the idea that the autonomy of human decision-making must be preserved during human-AI interactions, whether that choice is individual, or the freedom to choose together, such as the inviolability of democratic rights and values, also being linked to technological self-sufficiency of Nations/States;

- *Human Formation/Education:* such principles defend the idea that human formation and education must be prioritized in our technological advances. AI technologies require a considerable level of expertise to be produced and operated, and such knowledge should be accessible to all. This principle seems to be strongly tied to Labor Rights. The vast majority of documents concerned with workers and the work-life point to the need for re-educating and re-skilling the workforce as a mitigation strategy against technological unemployment;

- *Human-Centeredness/Alignment:* such principles advocate the idea that AI systems should be centered on and aligned with human values. AI technologies should be tailored to align with our values (e.g., value-sensitive design). This principle is also used as a "*catch-all*" category, many times being defined as a collection of "principles that are valued by humans" (e.g., freedom, privacy, non-discrimination, etc.);

- *Intellectual Property:* this principle seeks to ground the property rights over AI products and/or processes of knowledge generated by individuals, whether tangible or intangible;

- Justice/Equity/Fairness/Non-discrimination: this set of principles upholds the idea of non-discrimination and bias mitigation (discriminatory algorithmic biases AI systems can be subject to). It defends the idea that, regardless of the different sensitive attributes that may characterize an individual, all should be treated "*fairly*";

- *Labor Rights:* labor rights are legal and human rights related to the labor relations between workers and employers. In AI ethics, this principle emphasizes that workers' rights should be preserved regardless of whether labor relations are being mediated/augmented by AI technologies or not. One of the main preoccupations pointed out when this principle is presented is the mitigation of technological unemployment (e.g., through *Human Formation/Education*);

- *Open source/Fair Competition/Cooperation:* this set of principles advocates different means by which joint actions can be established and cultivated between AI stakeholders to achieve common goals. It also advocates for the free and open exchange of valuable AI assets (e.g., data, knowledge, patent rights, human resources) to mitigate possible AI/technology monopolies;

- *Privacy:* the idea of privacy can be defined as the individual's right to "*expose oneself voluntarily, and to the extent desired, to the world.*" In AI ethics, this principle upholds the right of a person to control the exposure and availability of personal information when mined as training data for AI systems. This principle is also related to concepts such as data minimization, anonymity, informed consent, and other data protection-related concepts;

- *Reliability/Safety/Security/Trustworthiness:* this set of principles upholds the idea that AI technologies should be reliable, in the sense that their use can be verifiably attested as safe and robust, promoting user trust and better acceptance of AI technologies;

- *Sustainability:* this principle can be understood as a form of "*intergenerational justice,*" where the well-being of future generations must also be counted during AI development. In AI ethics, sustainability refers to the idea that the development of AI technologies should be carried out with an awareness of their long-term implications, such as environmental costs and non-human life preservation/well-being;

- *Transparency/Explainability/Auditability:* this set of principles supports the idea that the use and development of AI technologies should be transparent for all interested stakeholders. Transparency can be related to "*the transparency of an organization*" or "*the transparency of an algorithm.*" This set of principles is also related to the idea that such information should be understandable to nonexperts, and when necessary, subject to be audited;

- Truthfulness: this principle upholds the idea that AI technologies must provide truthful information. It is also related to the idea that people should not be deceived when interacting with AI systems. This principle is strongly related to the mitigation of automated means of disinformation.

These 17 principles contemplate all of the values and worries we could interpret. This typology enabled us to encompass all normative principles, recommendations, tools, and general normative discourse expressed in our sample.

The first phase of our study also determined certain "*categories/types*" assigned to each document in the second phase. These types were determined by (1) the nature/content of the document, (2) the type of regulation that the document proposes, (3) the normative strength of this regulation, and (4) the impact scope that motivates the document.

The first type relates to the nature/content of the document:

- *Descriptive:* descriptive documents take the effort of presenting factual definitions related to AI technologies. These definitions serve to contextualize "*what we mean*" when we talk about AI, and how the vocabulary utilized in this field can be understood;

- *Normative:* normative documents present norms, ethical principles, recommendations, and imperative affirmations about what such technologies should, or should not, be used/developed for;

- *Practical:* practical documents present development tools to implement ethical principles and norms, be they qualitative (e.g., Self-assessment surveys) or quantitative (e.g., debasing algorithms for ML models).

These first three categories were defined as mutually inclusive, meaning that documents could be, for example, descriptive and normative, normative and practical, all three types, only one type, etc.

The second type relates to the form of regulation that the document proposes:

- *Government-Regulation:* this category is designed to encompass documents made by governmental institutions to regulate the use and development of AI, strictly (Legally binding horizontal regulations) or softly (Legally non-binding guidelines);

- *Self-Regulation/Voluntary Self-Commitment:* this category is designed to encompass documents made by private organizations and other bodies that defend a form of Self-Regulation governed by the AI industry itself. It also encompasses voluntary self-commitment made by independent organizations (NGOs, Professional Associations, etc.);

- *Recommendation:* this category is designed to encompass documents that only suggest possible forms of governance and ethical principles that should guide organizations seeking to use, develop, or regulate AI technologies.

We defined these categories as mutually exclusive, meaning that documents could only be one of the three established categories.

The third type relates to the normative strength of the regulation mechanism proposed by the document. For this, two categories were defined:[27]

- *Legally non-binding guidelines:* these documents propose an approach that intertwines AI principles with recommended practices for companies and other entities (i.e., soft law solutions);
- *Legally binding horizontal regulations:* these documents propose an approach that focuses on regulating specific uses of AI on legally binding horizontal regulations, like mandatory requirements and prohibitions.

We defined them as mutually inclusive, meaning that documents could present both forms of regulations.

The final type relates to the impact scope that motivates the document. With impact scope, we mean the dangers and negative prospects regarding the use of AI that inspired the type of regulation suggested by the document. For this, three final categories were defined and also posed as mutually exclusive:

- *Short-Termism:* this category is designed to encompass documents in which the scope of impact and preoccupation focus mainly on short-term problems, i.e., problems we are facing with current AI technologies (e.g., algorithmic discrimination, algorithmic opacity, privacy, legal accountability);
- *Long-Termism:* this category is designed to encompass documents in which the scope of impact and preoccupation focus mainly on long-term problems, i.e., problems we may come to face with future AI systems. Since such technologies are not yet a reality, such risks can be classified as hypothetical or, at best, uncertain (e.g., sentient AI, misaligned AGI, super intelligent AI, AI-related existential risks);
- *Short-Termism & Long-Termism:* this category is designed to encompass documents in which the scope of impact is both short and long-term, i.e., they present a *"mid-term"* scope of preoccupation. These documents address issues related to the Short-Termism category, while also pointing out the long-term impacts of our current AI adoption (e.g., AI interfering in democratic processes, autonomous weapons, existential risks, environmental sustainability, labor displacement, the need for updating our educational systems).

While in the first phase of our analysis, our team reviewed the entirety of our sample (each team member with its assigned quota), in phase two, a single team member reviewed all 200 documents. We concluded this approach would result in a standardized final sample. Thus, all post-processed documents passed the same criteria (and evaluator). In cases where uncertainties between classifications arose, we reached a consensus as a team.

We sought to establish our types and categories in the most objective way possible, as presented above. However, we recognize that even objective parameters are perceived and analyzed by subjective entities (people). Even if our final result possesses evaluation biases (some of our types may still be subject to interpretation and discussion), they all come from one source, consensually validated.

An abstract was written for each document in our sample as a final contribution. The reader can quickly scan the research contents thanks to the abstracts. We also attached their URL to each sample, which linked to the institution's website and any other significant attachments cited in the original document.

In the end, all documents received 13 features: Origin Country, World Region, Institution, Institution Type, Year of Publication, Principles, Principles Definition, Gender distribution, Size, Type I (Nature/Content), Type II (Form of Regulation), Type III (Normative Strength), Type IV (Impact Scope), plus Document Title, Abstract, Document URL, and Attachments.

We used all the information obtained during the second phase to create a database that feeds our visualization tool. The dashboard was created using the Power BI tool. A secondary dashboard (open-source) has been developed using the Dash library.[28]

The main difference between our tool from Hagendorff's table [21] and Fjeld et al. [15] graphs are its interactivity and the possibility to combine different filters without being restricted to preconfigured orderings. This enables researchers to draw on it to problematize characteristics found in their regions, map trends and behaviors, or investigate categories related to their particular research focus.

Another distinguishing feature of our tool is its ability to condense large amounts of information into a single visualization panel. Our choice for such a way of presenting our data was to make it easier to interpret how certain features interact with others. Previous works demonstrate the statistical distribution of certain features, such as country of origin and type of institution. But how are these features related to each other? For example:

- *Of the countries that have published the most papers, what are the main and least, advocated principles?*

- *What are the major concerns of NGOs?*

- *Which institutions are more concerned with the long term?*

- *Which governments are prioritizing stricter forms of regulation? And of these governments, what are the main principles advocated?*

- *Which institutions defend the least cited principles?*

- *How is the gender distribution among authors from governmental institutions?*

---

[27]Such categories were defined based on the definitions from the Innovative and Trustworthy AI [27].

[28]All resources are available in the following URL: https://nkluge-correa.github.io/worldwide_AI-ethics/.

The reader can access all this information by selecting and combining different categories on our panel.

## 4   Results

Our tool allows users to explore different statistics in varied forms. Users can select specific categories (e.g., specific countries or world regions) and see how the whole landscape of documents changes. The dashboards also present a detailed resume of each document (all of its typological classifications and information), as well as a page containing all paraphrased descriptions of gathered ethical principles (e.g., all accounts of "*Accountability/Liability*").

When we examine our sample through a "*country*" level of granularity, we see that the bulk (13 countries = 77%) of our total sample size is represented by the United States of America, the United Kingdom,[29] Germany, Canada, China, Japan, France, Finland, Netherlands, Switzerland, Belgium, Brazil, and South Korea, while a myriad of 24 countries (12,5%) represents the remainder of our sample, along with Intergovernmental organizations, like the EU (9 = 4,5%) and the UN (6 = 3%).

Looking at the distribution among world regions (Fig. **??**) (aggregated by continent), we see that the bulk of produced documents come from Europe,[30] North America,[31] and Asia,[32] while regions South America, Africa, and Oceania represent less than 4,5% of our entire sample size, with countries like Brazil (3 = 1,5%) spearheading the "rest" of our world sample (Latin America, 7 = 3,5%). If it was not for the significant participation of Intergovernmental Organizations, like NATO, UN, and UNESCO, which represent 6% of our sample size (13 documents), other world regions/countries would be even more underrepresented. However, this still excludes the Holy See/Vatican City and Palestine.[33]

Switching our gaze to institution types (Fig. **??**), except for institutions like IBM (5), Microsoft (4), and UNESCO (3), most other institutions do not have more than two published documents. We can also see that the bulk of our sample was produced by governmental institutions and private corporations (48%), followed by CSO/NGO (17%), non-profit organizations (16%), and academic institutions (12,5%). However, this trend only follows if we look at the totality of our sample size. If we look at documents produced by continents, for example, in North America (69), private corporations (24 = 34,7%) and non-profit organizations (18 = 26%) produced most documents, followed by governmental institutions (12 = 17,4%). Meanwhile, when we look at Europe, the global trend is restored.

An in-depth analysis segmented by countries shows that the engagement of particular AI stakeholders (i.e., institution types) differs between countries. For example, in China (11), the majority of documents are produced by academic institutions (5 = 45,4%), while in Germany (20), most documents in our sample came from private corporations (6 = 30%), and CSO/NGO (4 = 20%).

When we examined gender distribution among authors (removing documents with unspecified authors),[34] we performed an analysis based on the first names of each author. Given the variety/diversity that names can possess, it was necessary to use automation to predict gender encodes (male/female).

---

[29]In this study, the United Kingdom (England, Scotland, Wales, and Northern Ireland) has been considered as a country unit, even though technically this is not the case.

[30]Especially countries from Western Europe (63 = 31,5%), like the United Kingdom (24 = 12%) and Germany (20 = 10%).

[31]United States of America (58 = 29%) and Canada (11 = 5,5%), that together represent a third of our sample size.

[32]Mostly represented by East Asian countries (23 = 11.5%), like China (11 = 5,5%) and Japan (8 = 4%).

[33]Interestingly, the only document produced by a religious institution in our sample is the "*Rome Call for AI Ethics*," produced by the Pontifical Academy for Life (Vatican City). Meanwhile, we did not find a document produced by the Palestinian government.

[34]The final count shows that 66% of the sample (132 documents) does not identify the authors.

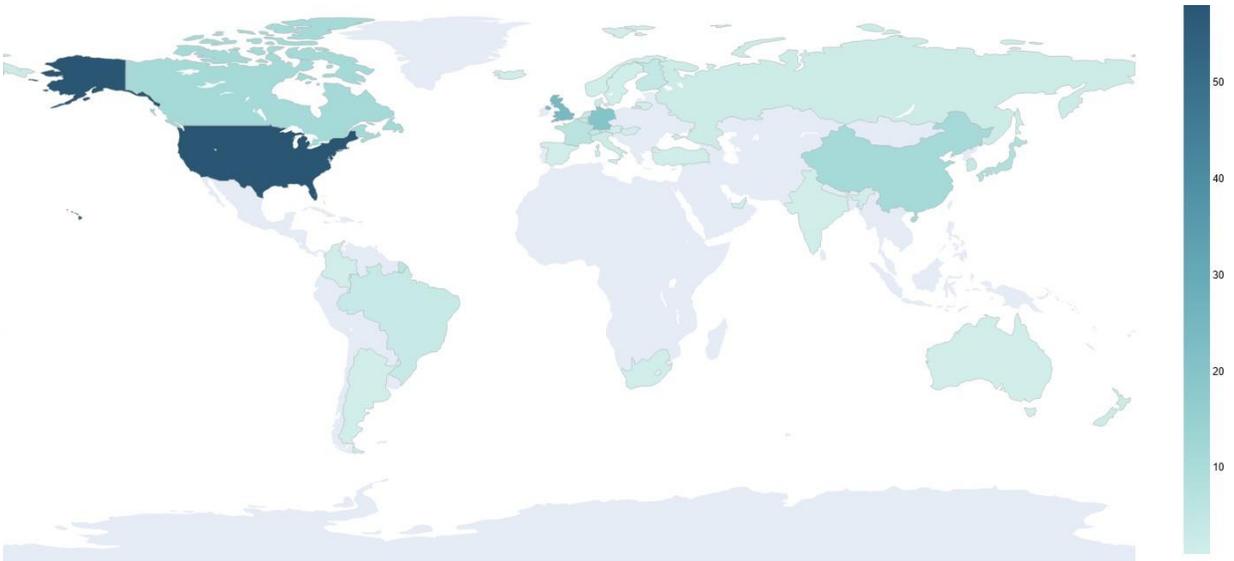

Figure 2: Number of published documents by country.

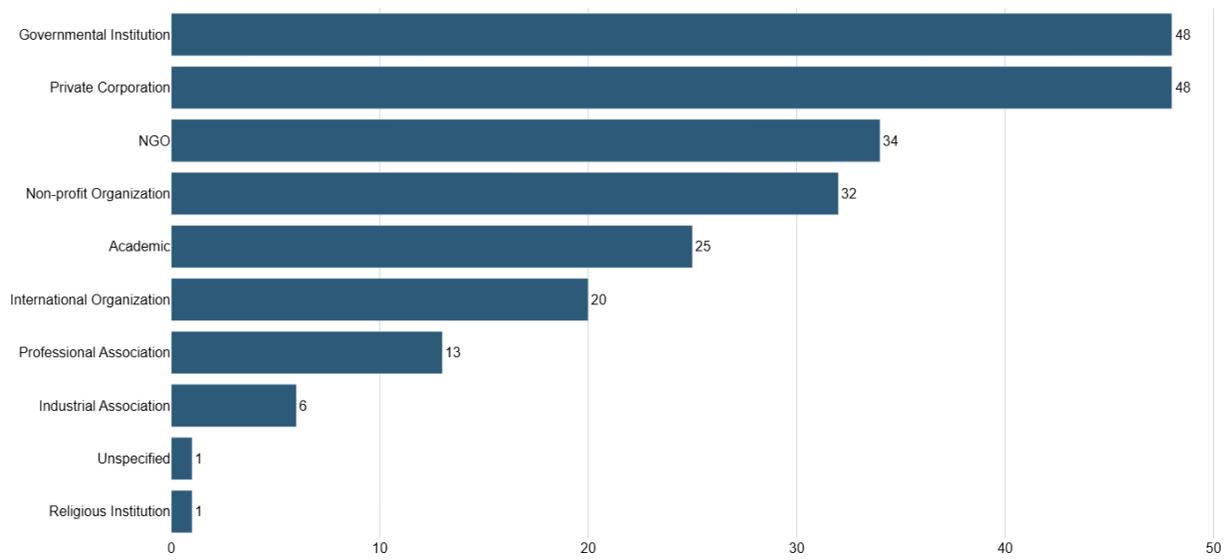

Figure 3: Publications by institution types.

To make an accurate inference, it was also necessary to extract (in addition to each author's name) the most likely nationality associated with each name. For this, we used (in addition to the country/origin of each document) nationalize.io,[35] an API service that predicts the nationality of a person given their first name. After that, we grouped the names of authors who had the same origin/nationality associated with their names. Finally, we used the API services of the genderize.io platform[36] to infer the gender of each name. We made each request by providing the name to be inferred and the ISO-2 code of the nationality associated with that name.

The result showed that the distribution of authors with "*male*" names was favorable in our database (549 = 66%). While 34% (281) of these names were inferred as "*female*."

Concerning the year of publication of the documents from our sample, one can see that the majority (129 = 64,5%) was published between 2017 and 2019. What we may call the "*AI ethics boom*" constitutes the significant production of documents in the year 2018, which represents 30,5% (61) of our entire sample (Fig. 4).

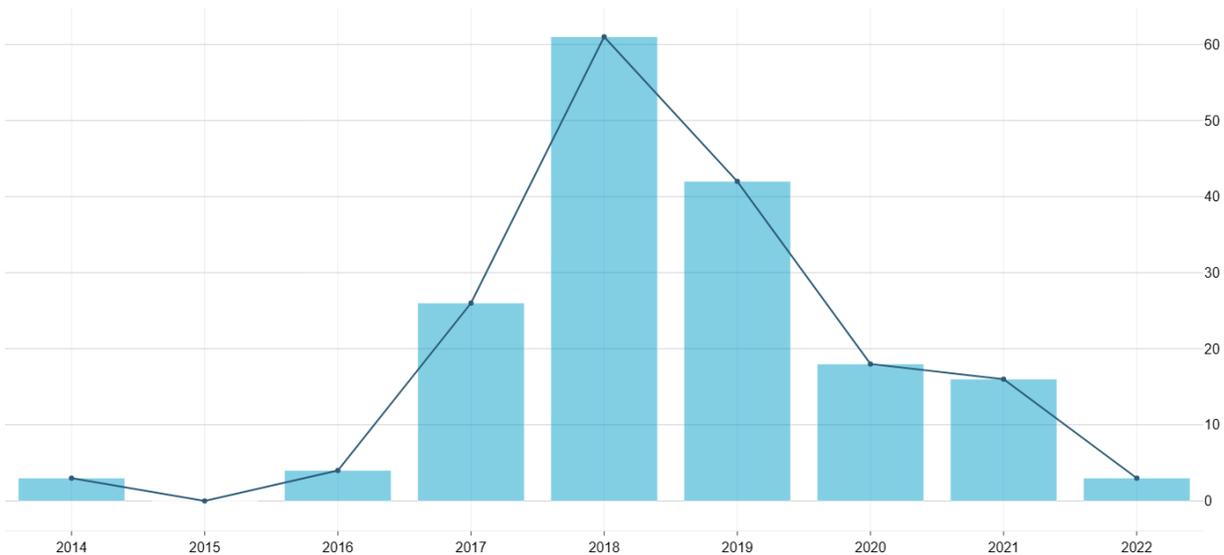

Figure 4: Nº of publications per year.

Regarding the previously defined typological categories, when looking at the document's Nature/Content, we found that the majority of our sample is from the normative type (96%), which a third of the time also presents descriptive contents (55,5%), and more rarely, practical implementations (2%).

When we look at the form of regulation proposed by the documents of our sample, more than half (56%) are only recommendations to different AI stakeholders, while 24% present self-regulatory/voluntary self-commitment style guidelines and only 20% propose a form of regulation administered by a given state/country.

This lack of convergence to a more "*government-based*" form of regulation is reflected in the normative strength of these documents, where the vast majority (98%) only serve as "*soft laws*," i.e., guidelines that do not entail any form of a legal obligation, while only 4,5% present stricter forms of regulation. Since only governmental institutions can create legally binding norms (other institutions lack this power), and they produced only 24% of our sample, some may argue that this imbalance lies in this fact. However, by filtering only the documents produced by governmental institutions, the disproportion does not go away, with only 18,7% of documents proposing legally binding forms of regulation. The countries on the front of this still weak trend are Canada, Germany, and the United Kingdom, with Australia, Norway, and the USA coming right behind.

Our last typology group is impact scope. Looking at the totality of our sample size, we see that short-term (47%) and "*mid-term*" (i.e., short-term & long-term = 52%) prevail over more long-term preoccupations (2%). When we filter our sample by impact scope and institution type, it seems to us that private corporations think more about the short-term (33%), governmental institutions about the short/long-term (28%), and academic (66%) and non-profit organizations (33%) with the long-term impacts of AI technologies.

Finally, examining the distribution of principles among our total sample size, we arrive at the following results: the top five principles advocated in the documents of our sample are similar to the results shown by Jobin et al. [31], and Hagendorff [21], with the addition of Reliability/Safety/Security/Trustworthiness (78%), which also was top five in Fjeld et al. [15] meta-analysis (80%) (Fig. 5).

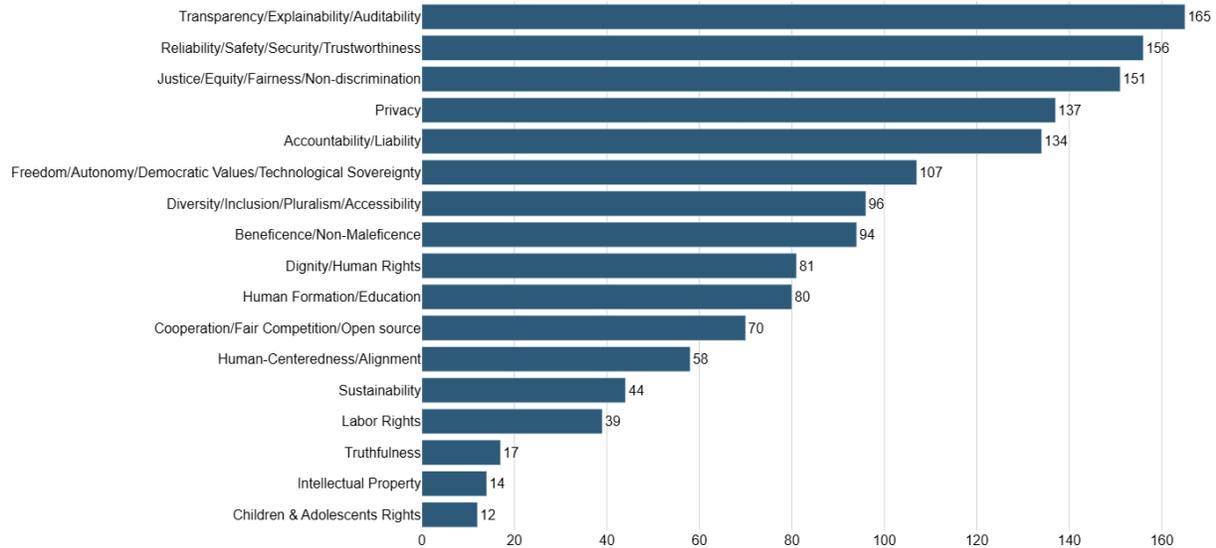

Figure 5: Number of times an aggregated principle was cited.

Looking at principle distribution filtered by continent, the top five principles remain the same in both North America and Europe, but the Asian continent introduces the principle of Beneficence/Non-Maleficence as is 5th (74%) most cited principle, putting Accountability/Liability in 6th place (70%). Filtering our results by country, we see no change in the top five principles when comparing EUA and the UK. However, looking under the top five principles, we begin to see differences, like Freedom/Autonomy/Democratic Values/Technological Sovereignty (38%) and Beneficence/Non-Maleficence (34,4%) being the 6th and 7th most cited principles in the EUA, and Open source/Fair Competition/Cooperation (45,8%) and Diversity/Inclusion/Pluralism/Accessibility (41,6%) being 6th and 7th most cited principles in the UK.

When examining principle distribution filtered by institution type, we also can find many insights. For example, looking at our total sample, we notice that the main preoccupation of governmental institutions (worldwide) is the need for transparent systems (89,5%), private corporations mainly advocate for Reliability (87,5%), and CSO/NGOs primarily defend the principle of fairness (88,2%).

Finally, regarding the "*principle definition divergence*," i.e., divergent forms of defining the observed ethical principles, we bring some of the cases that most sparked curiosity. First, let us look at our most cited principle: Transparency/Explainability/Auditability. When examining the definition proposed in "*ARCC: An Ethical Framework for Artificial Intelligence*" [65]:

- *"Promote algorithmic transparency and algorithmic audit, to achieve understandable and explainable AI systems. Explain the decisions assisted/made by AI systems when appropriate. Ensure individuals' right to know, and provide users with sufficient information concerning the AI system's purpose, function, limitation, and impact."*

In comparison with the one provided in "*A practical guide to Responsible Artificial Intelligence (AI)*" [49]:

- *"To instill trust in AI systems, people must be enabled to look under the hood at their underlying models, explore the data used to train them, expose the reasoning behind each decision, and provide coherent explanations* *to all stakeholders promptly. These explanations should be tailored to the different stakeholders, including regulators, data scientists, business sponsors, and end consumers."*

Both definitions seem similar, but the "*devil is in the details.*" Only the first definition entails the concept of auditing, which means (in some interpretations) a third-party review of the system in question. Also, while the first document mentions that "*one must explain*," "*ensure the right*," and "*provide enough information for people*," clearly implying the idea of a "*duty to explain*" (without specifying who should explain), coupled with the "*right to know*," the second document says that people have "*to be able to look under the hood*" (also without specifying who should be able to look), without bringing the idea of right or duty. Nevertheless, only the second one proposes that this knowledge should be tailored and accessible to different stakeholders since an explanation fit for a machine learning engineer may not be understandable to a consumer.

Keeping in mind that the concept of transparency is a well-fundamental idea/concept in AI (especially machine learning research), what kinds of differences may occur when we look at "*not so well defined*" principles, like human-centeredness? In "*Data, Responsibly (Vol. 1) Mirror, Mirror,*" [33], we find the following recommendation:

- *"Maybe what we need instead is to ground the design of AI systems in people. Using the data of the people, collected and deployed with an equitable methodology as determined by the people, to create technology that is beneficial for the people."*

While in "*Everyday Ethics for Artificial Intelligence,*" [11] following norm is suggested:

- *"AI should be designed to align with the norms and values of your user group in mind."*

The first document mentions ideas like "*the use of an equitable methodology*" and "*technology that is beneficial for the people.*" This idea of "*people*" seems to refer to a large and diverse group (perhaps "*all people*"). Meanwhile, the second specifically states "*your user group in mind,*" which could mean "*a small and select group of people,*" if that is what the designers have in mind as "*their users*."

Many other differences can be found in our sample, for example:

- "*Tieto's AI ethics guidelines*" [12] takes a different take on explainability, saying its systems "*can be explained and explain itself*", placing some of the responsibility on explainability in the AI system itself, making it a "*stakeholder*" in the accountability chain (a curious approach);

- The "*The Toronto Declaration*" [44] gives an extensive and non-exhaustive definition of what "*discrimination*" means under international laws, while most other documents resume themselves by only citing the concept, leaving open to interpretation the types of "*discrimination that is permissible*";

- In "*Artificial Intelligence and Machine Learning: Policy Paper,*" [63] fairness is related to the idea of "*AI provides socio-economic opportunities for all*" (benefits), in "*Trustworthy AI in Aotearoa: AI Principles*" [45] fairness is also defined as "*AI systems do not unjustly harm*" (impacts), which we can relate to the difference between certain notions of algorithmic fairness;

- While some documents (e.g., "*Telefónica´s Approach to the Responsible Use of AI*" [53] state how privacy and security are essential for AI systems developments, only a few define (e.g., "*Big Data, Artificial Intelligence, Machine Learning, and Data Protection*" [25] what "*good privacy criteria*" are (e.g., data minimization);

- And as a final example, while most documents interpret accountability/liability as "*developers being responsible for their projects*" (e.g., "*Declaration of Ethical Principles for AI in Latin America*" [35], some documents also put this responsibility on users, and even algorithms "*themselves*" (e.g., "*The Ethics of Code: Developing AI for Business with Five Core Principles*" [48]).

Besides the ones mentioned above, many other forms of analysis are possible, for example, the shift of attention between principles through the passing of the years, or more detailed "*multi-filtered*" forms of analysis, like:

- What is the principle most defended by private corporations worldwide? (Reliability).

- What is the principle that possesses the least amount of practical documents proposing solutions for their specific problems in North America? (Children & Adolescents Rights).

- Which country is at the front in terms of legally binding regulations aided by practical tools to assist AI developers? (United Kingdom).

All these results are available in our panel.

# 5   Discussion

The first point we would like to explore is the apparent unwavering distribution of documents into world regions/countries. Even with a sample size twice as large as the one analyzed by Jobin et al. [31], we seem unable to escape this result. We affirm that our results should be viewed as a "*true snapshot*" of the current AI Ethics global landscape. Other academic sources do provide some perspective on the subject, and, in the end, the pictures do not match. According to Savage [56], from 2016 to 2019: "*China's output of AI-related research increased by just over 120%, whereas output in the USA increased by almost 70%. In 2019 China published 102,161 AI-related papers, and the USA published 74,386.*".

Based on our analysis of the AI Index 2022 Annual Report [72], the top three countries by the Vibrancy Ranking[37] score are the US, China, and India. And with more than 52 billion USD invested in the AI field in 2021, it is comprehensible that almost a third of our sample size (58 documents) comes from the USA. But what about China and India? According to Zhang et al. [72], China has far surpassed the USA in terms of journal/conference publications and citations, while most of the "*AI talent concentration*" is found in India (followed by countries like the US, South Korea, Israel, and Singapore). However, China represents only 5,5% of our total sample (India being 0,5%).

Does this mean these places, like many others still underrepresented in our sample, are not paying attention to the current AI ethics debate? We believe this would be a wrong conclusion. We argue that a vibrant, unique, and culturally shaped discussion exists in places we are still unaware of, either due to language barriers or other factors.[38] For example, Kiemde and Kora [34], after acknowledging the dominance of Western values over the African population and the diaspora, where contributions to the debate remain very little (or not mainstream), bring an insightful discussion about the state-of-the-art on AI use in Africa (which is still controlled and managed by western monopolies), the foreign abuses suffered by the African population through AI experimentation [18], and the current state of affairs surrounding AI ethics and governance.

According to Kiemde and Kora [34], 17 of the 55 African Union (AU) member states possess data protection and privacy legislation, while Mauritius has announced the establishment of a National AI Council, also being the first African state to present an AI strategy/roadmap. The authors also demonstrate in their review a collection of published papers and documents about AI ethics in Africa [19, 54, 66, 20], and other underrepresented countries [5], which helps us to show that there is AI Ethics in Africa and probably in all other states that did not show up in our sample. They only do not come in the format we were first looking for.

However, the insights gathered from our sample are not entirely misleading. According to the AI Index of 2022 [72], private investment in AI reached an all-time high, surpassing the USD 90 billion mark, becoming more centralized (fewer companies funded/started, few companies receiving a greater partition of the total funding). This could help explain why so much of our sample (24%), tied with governmental institutions, comes from the private sector. Most AI technologies are created in the industry, and this industry, seeing the demands for regulation and accountability, quickly reacted by proposing its form of regulation: self-regulation (i.e., *we promise we will do well*). Many of such promises are, perhaps, genuine, but when governments and private institutions have "*the same weight*" in our sample, attention to the matter seems needed. The impasse between private and public interests remains a question that demands a proactive legislative push with measurable gauges of ethical practices beyond broad guidelines.

This fact may become more alarming when we look at the distribution of government documents that opt for "*soft*" forms of regulation (91,6%). The critique that "*ethical principles are not enough to govern the AI industry*" is not a new one [31, 37, 50, 21, 13]. However, some defend the "*principlism*" approach as something good [59]. A type of embryonic state that precedes stricter forms of regulation. Seger [59] argues that principles can work as a valuable starting point in discussions around regulation, helping us bring cultural context inside the formation of a new rule system.

Also, even considering that most countries in our sample[39] seem to opt for legally non-binding forms of regulation, there seems to be a growing adoption of stricter, and legally binding, solutions. The idea that "*Ethics*" and "*Compliance*" are separate domains seems to get ever-growing acknowledgment by countries such as Canada, Germany, and the United Kingdom (which comprise 66,6% of our total "*legally binding*" sample), while according to Zhang et al. [72], the legislative records on AI-related bills that passed into law grown from just one in 2016 to 18 in 2021, with Spain, the UK, and the USA being the top three "*AI-legislators*" from 2021, showing that in fact, we currently have more regulations related to AI then we ever had. Nevertheless, pushing or achieving legally binding regulations toward AI products does not end the ethical conundrum.[40]

---

[37]https://aiindex.stanford.edu/vibrancy/.

[38]Reminding that we focused on ethical guidelines, not academic papers.

[39]According to the document cited below, 14 countries of the European Union, those being Denmark, Belgium, the Czech Republic, Finland, France, Estonia, Ireland, Latvia, Luxembourg, the Netherlands, Poland, Portugal, Spain, and Sweden have stated favoritism towards soft law solutions [27].

Another factor worth contextualizing is the "*2018 AI Ethics boom*," i.e., the fact that almost a third of our sample (30,5%) got published in 2018 (64,5% if extended from 2017 to 2019). The AI Index report also points to this trend, where since 2014, we had a five-time increase in publications related to AI Ethics, where topics like algorithmic fairness have stopped being only academic objects of research and actual AI industry areas of R&D [72]. It is also interesting to see the shift of interest during the timeline we analyzed. In 2014, the top-cited principles were Fairness, Reliability, and Dignity (Transparency was not even in the top 10 at this time), and in 2016, Accountability, Beneficence, and Privacy received more attention (Accountability being the number one concern of documents published in 2017). But in 2018, Transparency (Explainable AI/XAI, Mechanistic Interpretability) became the dominant topic of concern.

What factors could explain this shift in attention? We can start by analyzing historical marks in those years that may be relevant to the field. AI research has seen a new wave of interest after its last winter since a deep neural network won the ImageNet 2012 Challenge, consider by many as the cementing of deep learning as the new parading in AI research. In subsequent years, we saw many remarkable feats that this type of system could perform, like a self-taught reinforcement learning agent that became the world's "*Best Go Player*" [60], and machine learning models that can create other machine learning models [74].

Deep learning systems rapidly became utilized in many real-world applications, making technology with little theoretical (but strong empirical support) backing widely adopted in many critical and sensitive areas of our society. Something that later would prove to be prone to lead to many unwanted consequences. Consequences may be a possible source of the sudden AI Ethics boom we experienced between 2017 and 2019.

Some high-profile cases that are worth mentioning are the COMPAS software use, which in the year 2016, Angwin et al. [4] showed that "*blacks are almost twice as likely as whites to be labeled a higher risk but not re-offend*." In 2018, we had the first case of a human killed by an Uber self-driving car [8]. And in the same year, the Cambridge Analytica case gained considerable media attention, where personal data was used without consent for personal profiling and targeting for political advertising [30]. We can also mention relevant works that helped cement the AI Ethics field as a popular area of research, like the celebrated book "*Weapons of Math Destruction*" [47]). All these events and many others helped bring life to a field that, according to the Google book n-gram viewer,[41] had almost no mention before 2012, and after 2014, saw a significant increase.

Perhaps some of these events could come to explain the swings of attention on AI Ethics. And perhaps transparency, or machine learning interpretability, assumed the top position of concern for being an older and more substantial area of study in machine learning [38], i.e., an area that could present some solid techniques and results for its problems. In a more critical sense, one could say that "*it is better to show concern with a problem that we have some idea on how to start solving (interpretability) than problems we have a little-to-none idea on how to approach (truthfulness and labor rights)*." As Hagendorff [21], asks: "*What does a 'human-centered' AI look like?*" (p. 8).

Meanwhile, we can see that only 55,5% of documents (111) seek to define what is the object of their discourse, i.e., "*we are talking about autonomous intelligent systems, and this is what we understand as an autonomous intelligent system*." This is a curious phenomenon, more so if we acknowledge that there is no consensual definition of what "*Artificial Intelligence*" is and what is not [39]. There are many interpretations and contesting definitions, which may prove to be a challenge for regulating organizations, i.e., there needs to be a rigorous definition of what we are trying to regulate. If you choose to define "*AI*" as only "*systems that can learn*" you will leave outside your scope of regulation an entire family of systems that do not learn (rule-based systems) but can still act "*intelligently*" and autonomously (e.g., UAVs - Unmanned aerial vehicles).

We would also like to point out the seemingly low attention given to the "*long-term impacts*" of AI (1,5%). Even though there is a considerable amount of work produced on AI safety[42] [6, 62, 61, 3, 24, 14, 32, 23], many times the terms "*safety*", "*alignment*," or "*human-level AI*" are generically dismissed as not serious, or as Stuart Russell would say: "*myths and moonshine*" [51]. Possible explanations for this fact could be: (a) the AI community does not find these problems real; (b) the AI community does not find these problems urgent; (c) the AI community thinks we have more urgent problems at hand; or even (d) that the AI community does not know about issues like "*alignment*" or "*corrigibility*." If we look at the distribution of papers submitted in the NeurIPS 2021,[43] approximately 2% were safety-related (e.g., AI safety, ML-Fairness, Privacy, Interpretability).

---

[40]Legislation may be topical and narrow in scope.

[41]https://books.google.com/ngrams.

[42]Alignment Research Center, Ought, and Redwood Research are examples of organizations (all being non-profit) focused on applied alignment research. https://alignment.org/. https://ought.org/. https://www.redwoodresearch.org/.

[43]https://nips.cc/Conferences/2021.

Finally, let us shift our attention from "*what is being said*" to "*who is saying*." There are significant gender differences between the male and female authors in our sample. The fact that 64% (128 documents) of our sample have unidentified authors may still mask this inequality. Academic institutions (62% male, 38% female) and non-profit organizations (65% male, 34% female) "*approximate*" gender parity but still fall short of the 1:1 parity line.

It is important to notice that it is hard to find present-day data about gender disparity since many organizations do not provide this information. More importantly, gender prediction methods still contain a "*gender bias error rate*" [67] and possible misclassification. Although the tools we used can be regarded as accurate tools [58, 73], gender predictions by name are blindsided for non-binary gender accounts and disregard self-declaration (e.g., non-binary, genderfluid, queer, or transgender).

With an understanding of the method's limitation to gender prediction, the ethical implications of gender encoding with data analysis, and AI gender bias analysis, one fundamental ongoing discussion in the AI community, we considered that the data contribute to turning the gaps and bias increasingly apparent. Therefore, below we bring some statistics to explain this kind of disproportion.

Going back to the AI Index report of 2022 [72], we see that, to our surprise, AI skills penetration, when filtered by gender, from the top 15 countries listed, rates of females are higher than those of males in India, Canada, South Korea, Australia, Finland, and Switzerland. However, this does not mirror the rest of the globe. For example, in the AI Now report of 2018, Whittaker et al. [68] showed that 80% of the professors at the world's leading universities (e.g., Stanford, Oxford) are male. According to the US National Center for Education Statistics (NCES),[44] between 2008 and 2017, women earned only 32% of undergraduate degrees in STEM (even given that more women graduate in the US than men, 60% more), and 18% of degrees in Computer Science. In the UK, women account for only 16% of the tech industry [26], and in Silicon Valley, the male/female proportion is 4:1 [9].

Google is one of the few big tech companies that make their internal demographic distribution publicly available [17], we can look at it and see that there still is a gender gap, especially if we filter gender by race. In their 2022 report, Google proudly stated that "*Black+ representation grown 2x faster than Googlers overall,*" however, men (62,5%) are still more hired than women (37,5%) at Google (globally), and leadership positions are still predominantly held by man (69,4%), while Black, Latin, and Native American women represent only 19,2% of their female "*Googlers.*" Although the data provided by the 200 documents in our meta-analysis allowed us to show a gap between male and female authorship distribution, we recognize there is a broader limitation of gender representation and gender bias discussion in AI, such as the inclusion of transgender and non-binary individuals, sexist data-algorithmic bias, discrimination, and so forth [7].

The underrepresentation of women and minority groups in technology sectors should be approached in a multidisciplinary and multimodal (numerical, textual, images, etc.) approach [40], including various levels of analysis such as historical, political, and technological. In addition to considering the issue of the presence of gender biases in IA systems, it is necessary to recognize the importance of diversity within the social and economic justice debate for gender parity [70]. We hope to start a conversation in a future meta-analysis on gender matters, which seems important to ethics in the field.

This work represents a mere fraction of what our true global landscape on this matter is. There is much data to be collected and typologies to be improved, and we hope to have helped lay the ground for future meta-analysts.

## 6 Conclusion

In this work, we sought to bring new data, insights, tools, typologies, and our interpretation/description of our sample. These descriptions sometimes went against other statistical data outside our original distribution and scope of focus. Thus, we sought to substantiate our interpretations with information beyond what we collected in our analysis. We believe this shows the, perhaps obvious, necessity to look "*outside of the data*" and contextualize it. To do so, we aimed to collect data from documents throughout the world, which are diverse and multi-faceted. Although this diversity is not fully represented in the data available, we argue that these 200 documents paint the picture of a world needing clear and enforceable rules for AI development.

From these analyses, it was possible to diagnose at least 17 groups of principles listed among the 200 guidelines analyzed, with the first six being present in more than 50% of the guidelines. This information certainly contributes as a guide for the discussions that are taking place on how to regulate artificial intelligence, indicating what objectives/minimum rights should be protected by future legislation. Therefore, we consider that we achieved the initial proposal of this research.

---

[44]https://nces.ed.gov/programs/digest/d18/tables/dt18_318.45.asp?current=yes.

The development space of systems and applications that affect so many people is still a "*black box*" for most. All stakeholders should be aware of the inner working of the processes that regulate and control their surroundings. The process to change our current deregulated AI industry may require a push for stronger government regulation, combined with a change in the culture of AI development. Many developers are already aware of the risks and flaws of these types of technologies. Flaws that endanger the most vulnerable among us.

We recognize the limitations that we, as researchers, face. This is reflected in our findings, where diversity, information about non-hegemonic countries, and broader gender and LGBTQIA+ representation remain deficient. AI governance is still a new and open field that demands greater attention and exploration, and this research sought to bring new tools to future explorers. As a final statement, we remind the reader that this is an ongoing project, and we hope to keep refining our sample.

## Acknowledgments

This research was funded by RAIES (*Rede de Inteligência Artificial Ética e Segura*). RAIES is project supported by FAPERGS (*Fundaço de Amparo à Pesquisa do Estado do Rio Grande do Sul*), and CNPq (*Conselho Nacional de Desenvolvimento Científico e Tecnológico*). Researchers also received funding from the DAAD (Deutscher Akademischer Austauschdienst).

## Author Contributions

This study involved a collaborative effort from a multidisciplinary team to collect documents, analyze statistical data, create data visualization tools, and present the results. N.K.C. contributed to the statistical analysis, dashboard creation, data collection, and writing. C.G. participated in data collection, writing, and dashboard creation. J.W.S., C.D.C., and C.B. contributed to data collection and writing. E.P.P., D.M., R.M., L.G., and E.T. were responsible for data collection. N.O. is the coordinator of the RAIES project and this research. All authors reviewed and approved the final version of the article.